\documentclass[smallextended]{svjour2}                    
\smartqed  
\usepackage{graphicx}
\usepackage{units}
\usepackage{bm,amsfonts}

\newcommand{\bsigma}{{\boldsymbol \sigma}} 

\usepackage{epsfig}
\usepackage{color}
\graphicspath{{src_images/}}

\begin{document}

\title{A Note on Automatic Kernel Carpentry for Atomistic Support of
  Continuum Stress}
\subtitle{}


\author{Manfred H. Ulz \and Sean J. Moran}


\institute{Manfred H. Ulz \at
  Institute for Strength of Materials,
  Graz University of Technology,
  Kopernikusgasse 24/I,
  8010 Graz,
  Austria\\
  \email{manfred.ulz@tugraz.at} 
  \and
  Sean J. Moran \at
  Institute for Language, Cognition and Computation,
  University of Edinburgh,
  Informatics Forum,
  10 Crichton Street,
  Edinburgh EH8 9AB,
  United Kingdom\\
  \email{sean.moran@ed.ac.uk}
}

\date{~~}

\maketitle

\begin{abstract}
Research within the field of multiscale modelling seeks, amongst other
questions, to reconcile atomistic scale interactions with
thermodynamical quantities (such as stress) on the continuum scale.
The estimation of stress at a continuum point on the atomistic scale
requires a pre-defined kernel function.  This kernel function derives
the stress at a continuum point by averaging the contribution from
atoms within a region surrounding the continuum point. Commonly the
kernel weight assignment is isotropic: an identical weight is assigned
to atoms at the same spatial distance, which is tantamount to a local
constant regression model. In this paper we employ a local linear
regression model and leverage the mechanism of automatic kernel
carpentry to allow for spatial averaging adaptive to the local
  distribution of atoms. As a result, different weights may be
assigned to atoms at the same spatial distance. This is of interest
for determining atomistic stress at stacking faults,
  interfaces or surfaces.  It is shown in this study that for
crystalline solids, although the local linear regression
model performs elegantly, the additional computational costs
  are not justified compared to the local constant regression model.
\end{abstract}


\section{Introduction}
\label{sec:introduction}

Atomistic to continuum coupling seeks to link atomistic extensive
quantities (for example mass, momentum, energy) to a continuum
quantity of interest (for example, stress, heat flux). This link has
previously been achieved by defining a space averaging volume (as
given by a \emph{kernel function}
\cite{Irving50,Noll55,Hardy82,Murdoch83}) over which the atomistic
extensive quantities are smoothed to compute the desired continuum
quantity. The kernel function adheres to the standard
  constraints of a proper smoothing function \cite[Section
    4.2]{Murdoch12}.

Recently, interest in modelling atomistic stress
\cite{Zimmerman04,Admal10} has been growing primarily due to the rise
in computational power and the development of atomistic to continuum
multiscale methods \cite{Curtin03,Miller09,Zeng10}. Furthermore,
attempts have been made to determine the kernel function using
statistical arguments \cite{Murdoch94,Murdoch12}, correlation
functions from statistical mechanics \cite{Ulz13} or data-driven methods from
machine learning \cite{UlzMoran12,UlzMoran13}.  A different
expression for the atomistic stress, for the particular study of
free surfaces, has been investigated \cite{Cheung91,Sun06} based upon a
force-area concept.  In previous work, the influence of the kernel
decays equally in all spatial directions with increasing distance from
the centre of the kernel. In contrast, little is currently known about
the effect of allowing the kernel weighting to \emph{vary} in different
dimensions according to the local distribution of the data.  The
possibility of a-priori defining such a kernel function
according to a given anisotropic distribution of matter is discussed
in \cite{Murdoch94,Murdoch12}.  In this paper we address this gap in
the literature by investigating a non-parametric kernel regression
method \cite{Fan96,Hastie09}, known as local linear regression (LLR),
that permits the weighting given by the kernel to adapt to the local
distribution (or density) of the data. In contrast to~\cite{Murdoch94,Murdoch12}
our method is entirely data-driven, with the model constructed directly from
the atomistic data itself.

In this work we take the observation made by~\cite{UlzMoran13} as our starting 
point, namely that the standard definition of virial stress is equivalent to a 
local constant regression (LCR) model, and investigate the benefit of extending 
the LCR model to permit the kernel weighting to \emph{vary} along different spatial 
directions. LCR approximates the regression
function by a (local) constant. A well-known issue with LCR is that boundary problems can detract 
from the global performance of the estimator: at the boundary of the predictor 
space the kernel neighbourhood is \emph{asymmetric}, which can cause a substantial
increase in the bias of the estimate. For example, a kernel at a point at the boundary will be truncated in its extent - those neighbouring data-points within this truncated extent will disproportionately influence the resulting predicted average by pushing or pulling the 
average above or below the value it would ordinarily assume if there were no boundary at that point. 
This is also true of irregular
or non-uniform interior regions. Both situations can frequently be found in crystalline 
solids: namely at stacking faults and the surfaces of flaws. This bias somewhat restricts application of the LCR estimator to values in the interior of the material. 

In this contribution we study the applicability of LLR to the modelling of 
atomistic stress, which is a known technique for reducing the bias significantly at the boundaries
at a modest cost in variance. Rather than making a local constant approximation, as for LCR,  the LLR model instead fits a linear regression line through the observations in the neighbourhood of each target data-point.  It 
can be shown that fitting a linear regression estimate at each target point is the same as computing the response by taking a weighted summation of surrounding data-points, weighted by an \emph{equivalent or effective kernel} 
which has been automatically adapted to the density of the samples. This phenomenon is known as \emph{automatic kernel carpentry} in the statistical literature~\cite[Section 6.1.1]{Hastie09}. 
In this paper we are particularly interested in applying LLR to materials that have surfaces and flaws in their lattice where the data-adaptive property of the equivalent kernel should be particularly noticeable.  Our main point of 
comparison will be the LCR model advocated by~\cite{UlzMoran13}. 

The organization of the paper is as follows.
Section~\ref{sec:stress_definitions} includes a brief review of the virial
and Hardy definitions of stress. Section~\ref{sec:llr} introduces the LLR
regression model and describes the formulation of the equivalent
kernel which we incorporate into the atomistic stress definition. The
performance of these kernels is tested in
Section~\ref{sec:examples}. Finally Section~\ref{sec:conclusion} discusses our
findings and provides suggestions for future work in this area.

\section{Atomistic stress}\label{sec:stress_definitions}

The Irving and Kirkwood \cite{Irving50,Noll55} formulation of stress
relates continuum quantities such as stress and heat flux to
microscopic kinematics and kinetics (for an elaborate discussion see
\cite{Admal10,Zimmerman04}).  Given their definition, we can derive
the virial stress $\bsigma_v$ at a continuum point ${\bf x}$ by
introducing a space averaging volume ${\Omega}({\bf x})$ centred at
point ${\bf x}$ and a uniform kernel $\psi_i = 1/\Omega({\bf x})$ that
has the dimension of inverse volume:

\begin{eqnarray}\label{eq:virial_stress}
  {\bsigma}_v({\bf x}) &=
  - \sum_{i=1}^N
  \psi_i
  {\bsigma}_i
  &=
  - \sum_{i=1}^N
  \psi_i
  \left(
    m_i {\bf u}_i \otimes {\bf u}_i +
    \frac{1}{2!} \sum_{j=1,j \neq i}^N
    {\bf f}_{ij} \otimes {\bf x}_{ij} + \ldots
  \right),
\end{eqnarray}
where $N$ is the number of atoms contained within the averaging
volume, $m_i$ denotes the mass of atom $i$, and ${\bf u}_i$ is the
relative velocity of the given atom to the mean velocity of the $N$
atoms. The position vector of atom $i$ is given by ${\bf x}_{i}$, with
${\bf x}_{ij} = {\bf x}_{i} - {\bf x}_{j}$ denoting the distance
between atoms $i$ and $j$, and ${\bf f}_{ij}$ specifying the force on
atom $i$ due to its pair interaction with atom $j$. The stress in
Equation~\ref{eq:virial_stress} may be equated to the virial theorem
\cite{Clausius70,Maxwell70,Maxwell74} by replacing ${\Omega}({\bf x})$
with the total volume of the system.

Based upon the Irving and Kirkwood procedure we can obtain the Hardy
definition \cite{Hardy82,Murdoch83} of stress by introducing the
kernel function $\psi_i=\psi(|{\bf x}-{\bf x}_i|)$:

\begin{eqnarray}\label{eq:Hardy_stress}
  {\bsigma}_h({\bf x}) &=
  - \sum_{i=1}^N
  \left( \psi_i m_i {\bf u}_i \otimes {\bf u}_i +
    \frac{1}{2!} \sum_{j=1,j \neq i}^N
    B_{ij} {\bf f}_{ij} \otimes {\bf x}_{ij}
    + \ldots \right). 
\end{eqnarray}
The kernel function $\psi_i$ defines the extent of the space averaging
volume surrounding a continuum point \cite[Section 4.2]{Murdoch12}.
The Hardy definition of stress introduces a bond function $B_{ij}({\bf
  x}) = \int_0^1 \psi(|{\bf x} - {\bf x}_i + \lambda {\bf
  x}_{ij}|)\,d \lambda$ between atoms $i$ and $j$.

\section{Local Linear Kernel Regression}
\label{sec:llr}

Kernel regression \cite{Fan96,Hastie09} models the regression
relationship between an explanatory variable X and a response variable
Y as:
\begin{equation}\label{eq:llr_1}
  Y_{i} = m(X_{i}) + \epsilon_{i},
\end{equation}
where $i=1 \ldots N$ and $m(\bullet)$ is the regression function, N is
the number of data points and $\epsilon_{i}$ is independently and
identically distributed zero mean noise. The goal of kernel regression
is to estimate the mean response curve $m$ in the regression
relationship.  If we assume $m$ is smooth, namely observations close
to $X$ contain information about the value of $m$ at $X$, then we can
use the $P$-term Taylor expansion of $m$ at $X$ (if $X$ is
  near to the sample $X_i$) to estimate the value of the function at
$X$:
\begin{equation}\label{eq:llr_2}
  m(X_{i}) \approx \underbrace{ m(X) }_{\beta_0} +
  \underbrace{ m'(X) }_{\beta_1} (X_i - X) +
  \underbrace{ \frac{1}{2!} m''(X) }_{\beta_2} (X_i - X)^2 + \ldots.
\end{equation}
The parameters $\beta_k$, with $k=0 \ldots P$, can be estimated by
solving the following least squares optimisation problem:
\begin{equation}\label{eq:llr_3}
\min\limits_{{\beta_k}} \sum_{i=1}^N \left( Y_i - \beta_0 - \beta_1 (X_i - X) - \beta_2 (X_i - X)^2 - \ldots \right)^2 K(|X_i - X|),
\end{equation}
where the kernel function $K$ gives nearby samples higher weight than
more distant samples. The kernel function is constrained to be
non-negative, symmetric and uni-modal.

The previous treatment can be generalised into a multivariate (in this
case 3-dimensional) problem by considering a dataset of quadruples
$({\bf x}_1,Y_1), \ldots ,$ $({\bf x}_N,Y_N)$ with ${\bf x}_i =
[x_i,y_i,z_i]^T$. In this case the least squares optimisation problem
(Equation~\ref{eq:llr_3}) may be stated as:
\begin{eqnarray}\label{eq:llr_4}
  \nonumber
  \min\limits_{{\boldsymbol \beta}_k} \sum_{i=1}^N \Big( Y_i &-& 
  \underbrace{ m ({\bf x}) }_{{\beta}_0} - 
  \underbrace{ \nabla m ({\bf x}) }_{{\boldsymbol \beta}_1} \cdot ({\bf x}_i - {\bf x}) \\&-& 
  ({\bf x}_i - {\bf x}) \cdot \underbrace{ \frac{1}{2!} {\cal H} m ({\bf x}) }_{{\boldsymbol \beta}_2} \cdot ({\bf x}_i - {\bf x}) - 
  \ldots \Big)^2
  K(|{\bf x}_i - {\bf x}|),
\end{eqnarray}
in which $\nabla (\bullet)$ and ${\cal H} (\bullet)$ give the gradient
and the Hessian, respectively. The symbol $(\cdot)$ gives the dot
product and $(\otimes)$ the dyadic product

We introduce Voigt's notation, for example, given a symmetric
$3\times3$-matrix ${\bf A}$ we represent it as a vector
$\textrm{vn}\{{\bf A}\} =
[A_{11},A_{22},A_{33},A_{23},A_{13},A_{12}]^T$. By doing so we can
rewrite Equation~\ref{eq:llr_4} in matrix form as \cite{Hastie09,Fan96,Kay93}
\begin{equation}\label{eq:llr_5}
  \widehat{\bf s} = \arg \min\limits_{{\bf s}}
  ({\bf Y} - {\bf B} {\bf s})^T {\bf W} ({\bf Y} - {\bf B} {\bf s})
\end{equation}
with ${\bf Y} = [Y_1,Y_2,Y_3,\ldots]^T$, ${\bf s} = [\beta_0,
  {{\boldsymbol \beta}_1}^T, \textrm{vn} \{ {\boldsymbol \beta}_2 \}^T
  ,\ldots]^T$, ${\bf W} = \textrm{diag}[K(|{\bf x}_1 - {\bf
    x}|),K(|{\bf x}_2 - {\bf x}|),K(|{\bf x}_3 - {\bf x}|),\ldots]$
with ``$\textrm{diag}$'' defining a diagonal matrix.  Let ${\bf B}$ be
the regression matrix with the $i$th row being ${\bf b}_i$, where
vector ${\bf b}_i = [1, ({\bf x}_i - {\bf x})^T, \textrm{vn} \{ ({\bf
    x}_i - {\bf x})\otimes({\bf x}_i - {\bf x}) \}^T,\ldots]$.

For any given order $P$, the parameter ${\beta}_0$ provides the
estimated function value $m({\bf X})$ for a given observation ${\bf
  X}$.  We introduce a column vector ${\bf e}_1$ with its first
element equal to one and the rest to zero so as to extract ${\beta}_0$
from the least squares minimization problem (Equation~\ref{eq:llr_5}):
\begin{equation}\label{eq:llr_6}
  \widehat{m({\bf X})} = \widehat{{\beta}_0} =
  {\bf e}_1^T ({\bf B}^T {\bf W} {\bf B} )^{-1}
          {\bf B}^T {\bf W} {\bf Y} \doteq \sum_{i=1}^N l({\bf X}) Y_i.
\end{equation}
The matrix $( {\bf B}^T {\bf W} {\bf B} )^{-1}{\bf B}^T {\bf W}$ is
the Moore-Penrose inverse and $l({\bf X})$ is the so called
\emph{equivalent kernel} of local regression analysis.  The parameter $P$ determines which polynomial order is used
to locally approximate $m({\bf X})$: with $P=0$, local constant
regression is revealed, with $P=1$ we obtain local linear regression
and with $P > 1$ we obtain polynomials of increasing order.  Local
constant regression assumes the following form for ${m({\bf X})}$:
\begin{equation}\label{eq:llr_7}
  \widehat{m({\bf X})} = \frac{ \sum\limits_{i=1}^N K(|{\bf x}_i - {\bf x}|) Y_i } {\sum\limits_{j=1}^N K(|{\bf x}_j - {\bf x}|)} =
  \sum_{i=1}^N \psi(|{\bf x}_i - {\bf x}|) Y_i.
\end{equation}
Equation~\ref{eq:llr_7} coincides with the definition of the virial stress in
Equation~\ref{eq:virial_stress}. Firstly, by choosing $\psi$ to be a uniform
kernel function. And secondly, by setting $Y_i$ to a component of
${\boldsymbol \sigma}_i$ divided by $V_i={\cal V}/{\cal N}$ (with
${\cal V}$ being the volume of the entire system occupied by ${\cal
  N}$ atoms) forming a stress per atom value. In our experimental 
validation (Section \ref{sec:examples}) we restrict our attention to the LCR ($P=0$) and LLR
($P=1$) models.

\begin{remark}
  \label{remark:problem1}
  The equivalent kernel due to LLR may take negative values. Most
  previous related work \cite{Hardy82,Admal10} in the literature use a
  non-negative kernel function. Other references
  \cite{WebbIII08,Murdoch07} state the required properties of the
  kernel function without explicitly discussing if its either positive
  or negative. However, the kernel function may be allowed to take
  negative values in general \cite{Murdoch94,Murdoch12}.
\end{remark}

\begin{remark}
  \label{remark:problem2}
  The derivation of Hardy stress requires the kernel function to be
  solely dependent on a single distance from the continuum point at
  ${\bf x}$ to an atom under consideration at ${\bf x}_i$:
  $\psi_i=\psi({\bf r}_i)$ with ${\bf r}_i = {\bf x}_i - {\bf
    x}$. This allows for the desired property $\partial_{\bf
    x}\psi({\bf r}_i) = - \partial_{{\bf r}_i}\psi({\bf r}_i)$
  \cite{WebbIII08}.  Given this, an implementation of LLR with respect
  to Hardy stress is not straightforward. The equivalent kernel is
  dependent on both ${\bf r}_i$ and the distances to the neighbors
  ${\bf r}_j$: $l_i({\bf r}_i,{\bf r}_j)$. To circumvent this problem,
  the assumption can be made that only the distance ${\bf r}_i$ is
  subject of change, if ${\bf x}$ changes. This can be written as
  $\partial_{\bf x} l({\bf r}_i, {\bf r}_j) \bigm|_{{\bf r}_j=const.,j
    \neq i} \equiv - \partial_{{\bf r}_i} l({\bf r}_i,{\bf
    r}_j)$. Therefore, only the distance to atom $i$ may change, if
  the position of ${\bf x}$ changes. The distance to the remaining
  atoms cannot change. This is one reason, in addition to
  significantly higher computational costs \cite{UlzMoran13}, as to
  why we focus on the virial stress and do not elaborate
  on the investigation of Hardy stress.
\end{remark}

\section{Numerical examples}
\label{sec:examples}


We evaluate the local linear regression (LLR) model on two classical
examples of elasticity. Our first example is the distribution of
residual stress about the core of an edge dislocation in an elastic
plane solid. Secondly, we consider an infinite plate with a circular
notch under uniaxial load.  In both numerical experiments we
investigate a single crystal of copper in a face centred cubic lattice
using LAMMPS \cite{Lammps}. This model has $1000\times1000\times3$
unit cells (with length $a=3.615\unit{\AA}$) with periodic boundaries
on the two square surfaces and free boundaries on the remaining four
surfaces. An EAM potential is used with a cut-off radius of 2.5$a$
\cite{Zhou01}. Subsequently, energy minimization at zero temperature
with relative tolerance $10^{-15}$ guides the system to
equilibrium. To model plane strain, the atomistic system is reduced to
two dimensions in the $x,y$-plane by taking a unit cell slice of
thickness $a$ in the dimension bearing the periodic
boundaries. Furthermore, we constrain our space averaging volume to
have a circular cross section in the $x,y$-plane.

\subsection{Edge dislocation}
\label{subsec:example_a}

\begin{figure}[ht]
  \centering
  (a)\includegraphics[width=0.29\textwidth]{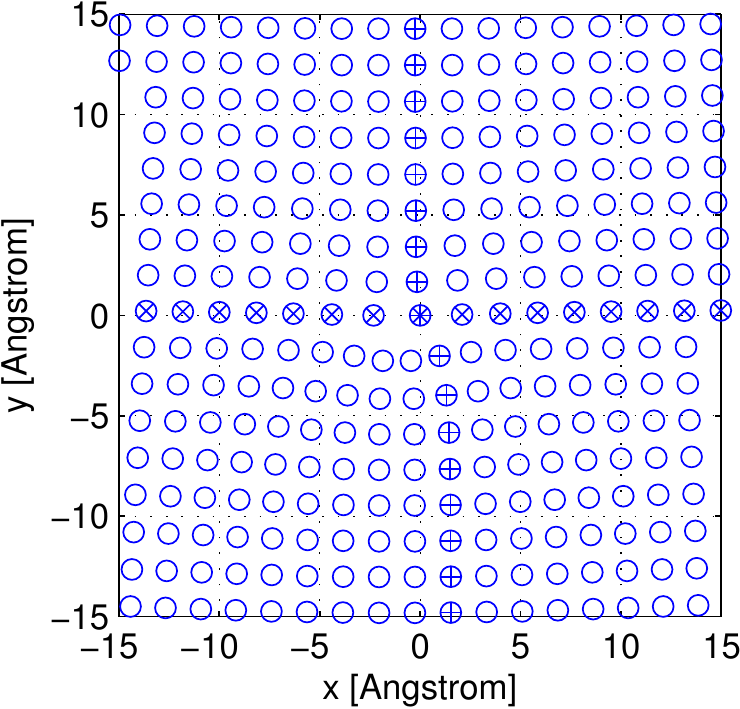}
  (b)\includegraphics[width=0.29\textwidth]{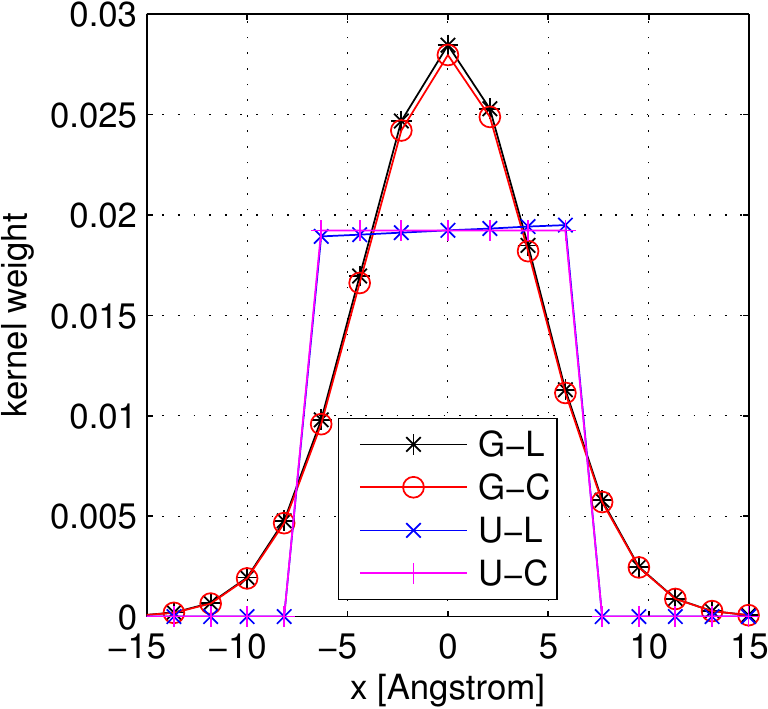}
  (c)\includegraphics[width=0.29\textwidth]{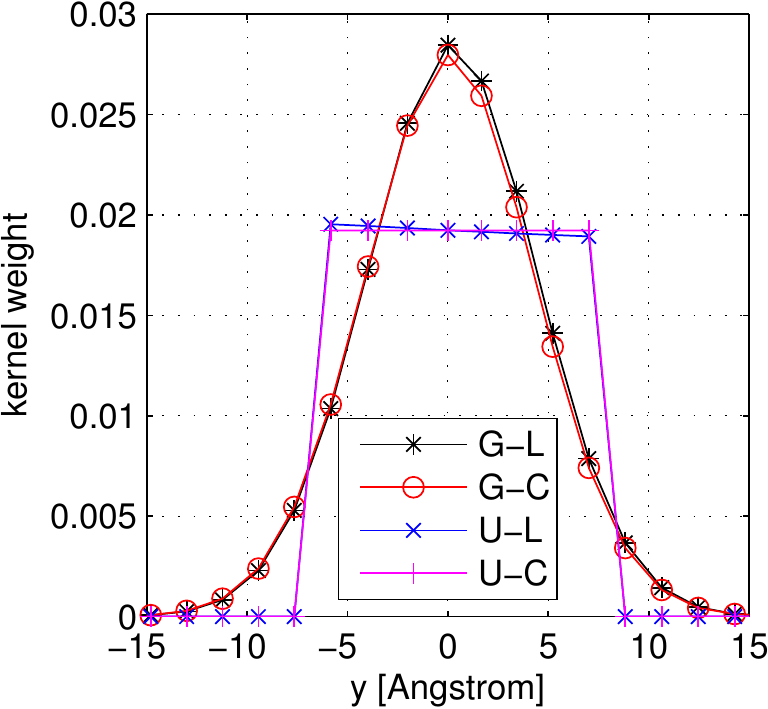}
  \caption{(a) The arrangement of the atoms in an edge dislocation
    gives a rather regular sample distribution. Lines of atoms in both
    x- ($\otimes$) and y-direction ($\oplus$) are selected. A Gaussian
    (G) and a uniform (U) kernel function for LCR (C) and LLR (L) are
    directly placed above the crossing atom of these two lines.  This
    atom is the centre of the shown coordinate system. Horizontal (b)
    and vertical (c) slices of the kernel functions are given for the
    selected atoms.}
  \label{fig:edge_dislocation_weights}
\end{figure}

In this experiment we take the base system and create a $\langle 100
\rangle$ edge dislocation at the centre of the simulation box before
energy minimization; this results in a Burgers vector oriented along
the $x$-direction. A circular slice is removed from the atomistic
system surrounding the origin of the coordinate system with a radius
of $20a$ in the $x,y$-plane and a thickness of $a$. This yields 5,024
data points, the spatial location of which can be seen in
Figure~\ref{fig:edge_dislocation_weights}.

To ascertain the effect of local constant regression ($P=0$ in
Equation~\ref{eq:llr_6}) and local linear regression ($P=1$ in
Equation~\ref{eq:llr_6}) on our atomistic dataset we compute the difference
between the kernel weights arising from both regression models. For
ease of exposition we select a Gaussian (with a standard deviation of
$4.31\unit{\AA}$) and a uniform kernel (with a bandwidth of
$7.50\unit{\AA}$ to adjust for the different shape of the kernel)
\cite{Marron88} for smoothing.  The kernels are placed over a
selected atom as shown in Figure~\ref{fig:edge_dislocation_weights} and the
associated weights are computed (slices of the kernel weights are
plotted in Figure~\ref{fig:edge_dislocation_weights}).  As can be observed,
the kernel weights arising from both regression models are effectively
identical. The reason for this is due to the almost uniform spacing of
the atoms which has the effect of rendering
the off diagonal blocks in ${\bf B}^T {\bf W} {\bf B}$ of
Equation~\ref{eq:llr_6} with values very close to zero in the LLR model. Therefore, the kernels for $P=0$ and $P=1$ are
practically identical (see \cite[Section II.D]{Takeda07} for a
  discussion).

We take the definition of virial stress from Equation~\ref{eq:virial_stress}
and replace the uniform kernel with the LLR equivalent kernel
according to Equation~\ref{eq:llr_6} and set ${\boldsymbol \sigma}_i
\rightarrow {\boldsymbol \sigma}_i/V_i$. With respect to the
analytical solution of anisotropic linear elasticity \cite{Hirth82}
(elastic constants taken from \cite{Lazarus49}), the LLR model obtains
a $0.126$ percent increase in root mean squared error (RMSE) as compared to the LCR model of
virial stress. This difference is negligible and indicates that there
is no substantial difference between the predicted stress arising from
models. Replacing the uniform kernel with a Gaussian kernel, thereby
yielding an altered virial stress, results in a similar observation:
the difference between the output of both models is barely perceptible
(relative increase in RMSE: $0.056$ percent).

\subsection{Infinite plate with circular notch}
\label{subsec:example_b}

In our second simulation we consider the problem of an infinite plate
with a circular notch subject to uniaxial loading. The base system is
altered by firstly removing a hole of radius $60a$ from the centre
following by the application of a uniform tensile traction of
$p=1\unit{GPa}$ on the edges normal to the $y$-direction. The
equilibrium state is found at a temperature of zero Kelvin by energy
minimization with relative tolerance $10^{-15}$.  An atomistic
solution to this problem was previously given in \cite{Admal10,Ulz13},
while an analytical solution of elasticity theory for the case of
plane strain was provided in \cite{Lekhnitskii63}.  However, the
effect of surface free energy is neglected in the latter as it is
considered to be small compared to the bulk energy in continuum
mechanics. In Section \ref{subsec:example_a}, based on a direct comparison of kernel
weights, we found that local linear
regression (LLR) is effectively identical to local constant regression
(LCR) when applied to an atomistic system that conforms to a regular grid. In
this experiment we compare both models in the situation where we have a 
pore in the material. This pore introduces a boundary in the system only 
one side of which contains atoms.

\begin{figure}[ht]
  \centering
  (a)\includegraphics[width=0.29\textwidth]{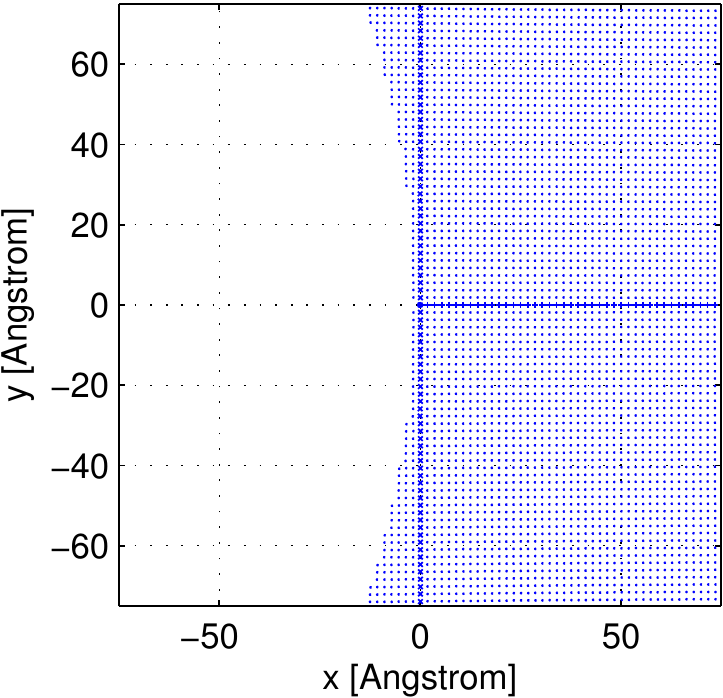}
  (b)\includegraphics[width=0.29\textwidth]{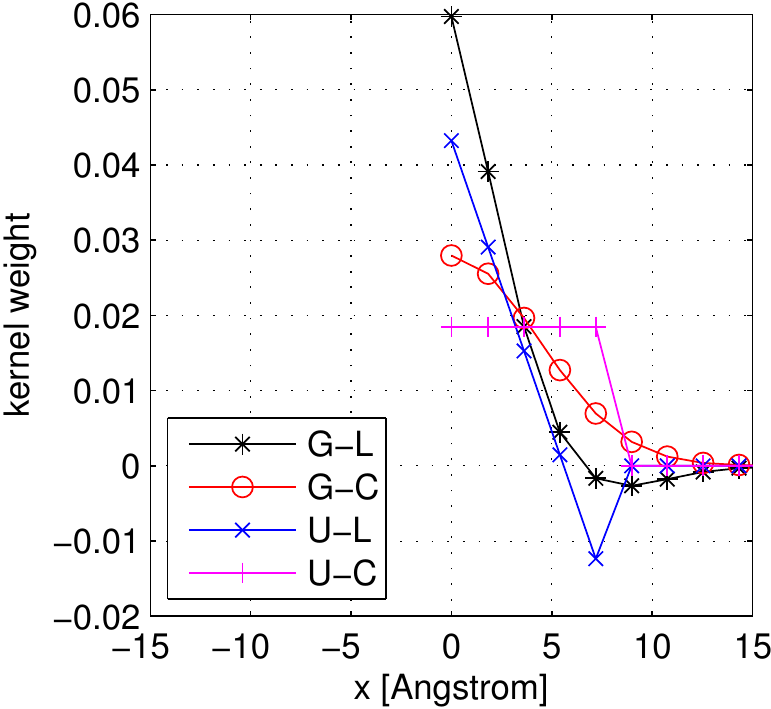}
  (c)\includegraphics[width=0.29\textwidth]{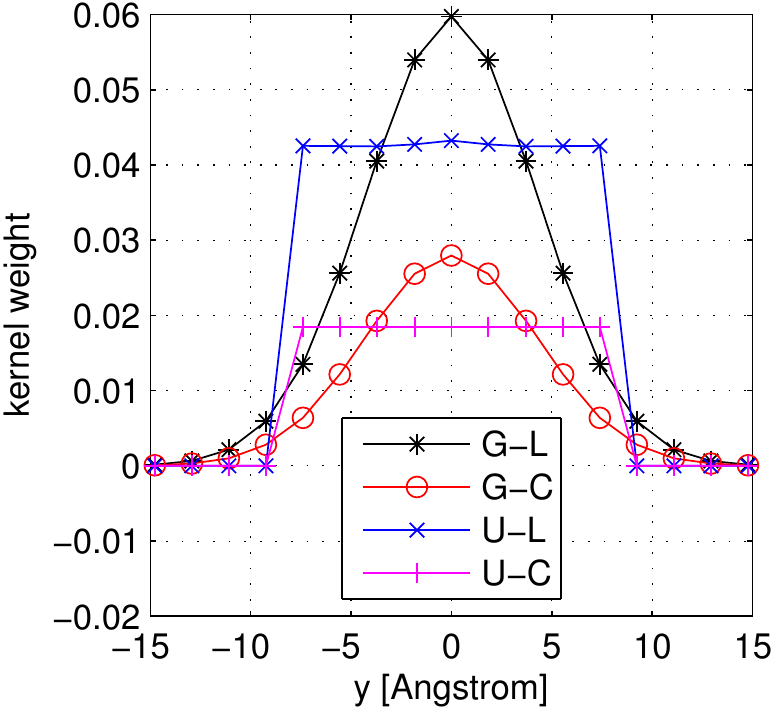}
  \caption{(a) The arrangement of the atoms at the notch. Lines of
    atoms in both x- ($+$) and y-direction ($\times$) are selected. A Gaussian
    (G) and a uniform (U) kernel function for LCR (C) and LLR (L) are
    directly placed above the crossing atom of these two lines. This
    atom is the centre of the shown coordinate system. Horizontal (b)
    and vertical (c) slices of the kernel functions are given for the
    selected atoms.}
  \label{fig:kirsch_problem_weights}
\end{figure}

We construct the simulation as follows: a circular slice with thickness $a$ centred at $x=60a$ and
$y=0\unit{\AA}$ is taken as our atomistic dataset. The slice has
radius $30a$ giving 6,329 data points
(Figure~\ref{fig:kirsch_problem_weights}). We fit both the LCR and LLR
models to this dataset. The assigned weights for both LCR and LLR are
shown in Figure~\ref{fig:kirsch_problem_weights} for a uniform and a
Gaussian kernel (as described in Section~\ref{subsec:example_a}). If we
take the Gaussian kernel as our example, we can immediately 
observe the effect of the data-adaptive equivalent kernel at the
boundary of the material (Figure~\ref{fig:kirsch_problem_weights}(b)). 
As there are no atoms  in the negative x-direction the kernels 
are therefore truncated in their extent in this region. The red line with circular 
markers denotes the Gaussian kernel weights in the LCR model, 
while the black line with asterisk-style markers denotes the 
equivalent kernel in the LLR model. It is clear that the equivalent 
kernel has adapted to the local density of the data at the boundary: versus
the LCR Gaussian kernel, the
equivalent kernel distributes substantially more weight to the cluster of
atoms lying on or very close to the boundary ($\le 2.5\unit{\AA}$), while
considerably downweighting those atoms located further away ($> 2.5\unit{\AA}$). The equivalent
kernel therefore compensates for the lack of atoms in the negative x-direction by up-weighting the contribution of the boundary atoms to
the estimation of the average stress at the boundary. As the stress at the boundary atoms is 
substantially different to that in the interior, this adaptation is beneficial for prediction of stress at the boundary - this estimate will be dominated by the stress of boundary atoms, with very 
little contribution from more distant atoms. The net result of this kernel 
adaptation is a better
fit to the atomistic data in the boundary region.

\begin{figure}[ht]
  \centering
  (a)\includegraphics[width=0.7\textwidth]{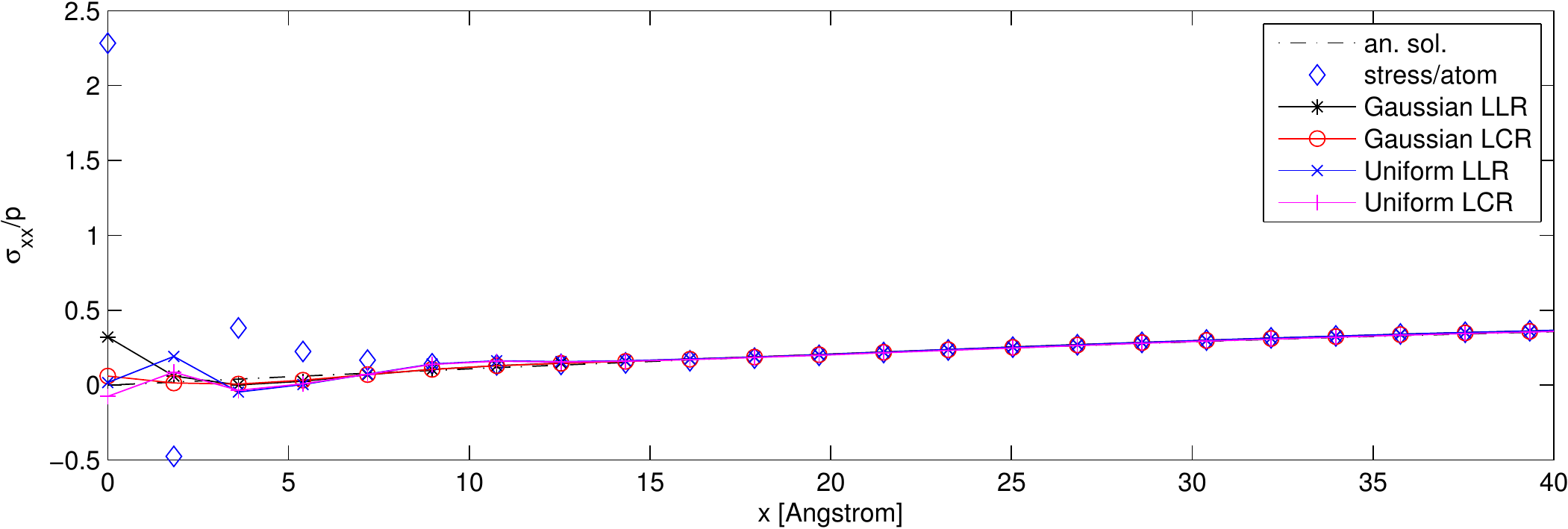}\\
  (b)\includegraphics[width=0.7\textwidth]{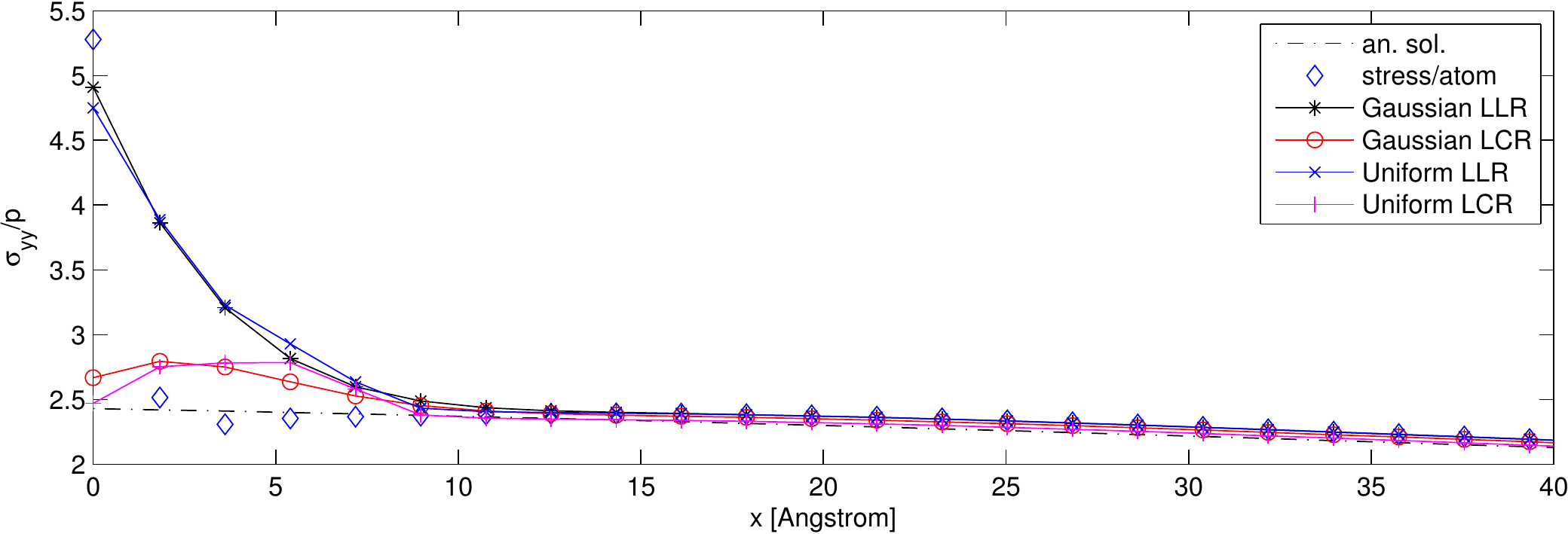}
  \caption{Stress distribution for notch problem.  Normalized normal
    stresses (a) $\sigma_{xx}$ and (b) $\sigma_{yy}$ in positive
    $x$-direction starting at the radius of the notch and at
    $y=$0\AA. The analytical solution (surface effects disregarded) is
    given by the dotdashed line, the volume-related stress per atom
    values (${\boldsymbol \sigma}_i/V_i$) are given by ($\Diamond$).}
  \label{fig:kirsch_problem_stress}
\end{figure}

Unfortunately, despite our expectations to the contrary, the improved fit at the boundary does not translate into an improved estimate of virial stress as given by Equation~\ref{eq:virial_stress}. The central cause of this are 
\emph{surface effects} which have an impact
on the atomistic stress per atom values ${\bsigma}_i$ given in
Equation~\ref{eq:virial_stress}.  More specifically, the surface effects cause the stress per atom value of the boundary atoms to be very different from the stress per atom values of those atoms only a few $\unit{\AA}$ into the material 
from the boundary. This can be seen by observing the patterns of the stress per atom values (indicated by the diamond symbol) from $0 \unit{\AA}$ to $10 \unit{\AA}$ in Figure~\ref{fig:kirsch_problem_stress}(b). Due to the large 
discontinuity in stress over a small distance in this region, the improved fit of the LLR model at the boundary translates into a much poorer fit in the region several units of 
$\unit{\AA}$ into the material. The LLR model effectively attempts to smooth out this discontinuity, resulting in a poor fit in the locality of the discontinuity. 
Ideally the impact of the surface-affected atoms should be restricted to the surface itself, necessitating that the kernel assign negligible weight to such atoms when computing the stress at points in the interior. However the 
equivalent kernel, due to its data-adaptive capability, assigns a much higher weight to boundary atoms than the LCR kernel, when computing the average stress for atoms in the region adjacent to the boundary ($0 \unit{\AA}$ 
to $10 \unit{\AA}$ in Figure~\ref{fig:kirsch_problem_stress}(b)). 
In contrast, the weights assigned to the surface atoms by the LCR kernel do not differ substantially to the weights assigned by that kernel to atoms in the interior, effectively curtailing the impact of surface effects on the predicted average stress of interior atoms. 
The net result is that the classical virial stress (that is, the LCR model), due to its poor fit at the boundary, actually yields a superior overall fit to the atomistic data due to its insensitivity to the stress discontinuity near the surface. This observation
negates any potential benefit brought about through modelling the dataset, and in particular, the boundary region, using LLR.

\section{Discussion and conclusion}
\label{sec:conclusion}

The objective of this paper was to investigate the applicability of
\emph{automatic kernel carpentry} for the purposes of modelling atomistic
stress using non-parametric kernel regression. In many materials of interest,
flaws or boundaries result in an asymmetry in the distribution of atoms. Automatic kernel
carpentry provides a principled means of adapting the weighting of the kernel according to the local density of the data samples. It is well known that the local linear regression (LLR) model yields a reduction in the bias 
of the regression function, with a modest increase in variance, yielding, in many cases, an overall improved fit to the data. This data-adapted kernel is known as the \emph{equivalent kernel} in the statistical literature. 
In this work we replaced the uniform kernel in the virial
stress by the equivalent kernel formed by a LLR model yielding an altered virial stress. The predictive accuracy of our LLR model was then compared to the local constant regression (LCR) model of~\cite{UlzMoran13}. The key finding of our experimental validation is that automatic kernel carpentry, as given by fitting an LLR model to atomistic stress, does not yield an increase in the accuracy of the 
virial stress estimate versus the LCR model of~\cite{UlzMoran13}.  The main barriers to
effective application of the LLR model proved to be both the highly
regular arrangement of atoms in the crystalline solids under study and the
surface-affected atoms at the boundary of the materials. 

Firstly, the
regular spacing of the atoms in crystalline solids resulted in the LLR
equivalent kernel to be effectively identical to that obtained using the LCR model in the interior. If the
atoms are placed in a perfect lattice there is no asymmetry in the
distribution of the atoms, and therefore no scope for kernel adaptation as all points in the space have equivalent density. The net effect of this is that the LCR model provides similar stress estimates to the LLR model, at a 
fraction of the computational cost. 
Secondly
surface effects, which cause the stress of surface atoms 
to vary greatly from adjacent atoms located in the inner regions of
the material, negated any benefit arising
from the boundary bias correcting capability of the LLR model. There is effectively a discontinuity in the stress distribution in the region near and at the surface, which contradicts the basic smoothness assumption of non-parametric kernel regression. In our experimental validation we observed that the LCR model is less effected by this discontinuity than the LLR model. The influence of
the highly surface-affected atoms at the boundary onto the averaged stress in the
proximity of the boundary is undesirably magnified by the weight distribution of the equivalent kernel, that is, substantially more weight is assigned to surface atoms versus those atoms in the interior of the material. In contrast, the averaged
stress given by the LCR model in the proximity of the surface is much less
influenced by surface effects. We identified the cause of this to be the more equal weighting applied by the LCR kernel to
atoms in the interior and to those at the surface, thereby providing an average that is less sensitive to discontinuities in the atomistic data. In other words, the LCR model, due to its increased bias at boundary regions, essentially restricts the contribution of surface atoms on the average stress prediction at points in the interior.
A local linear regression model with change points \cite{SanchezBorrego06} would be a possible means of handling these discontinuities in the atomistic data - we leave this investigation to future work.
Given these observations we found that
the additional computational effort of implementing the LLR model versus the
LCR model is not justified for the modelling of atomistic stress in crystalline solids.

Nevertheless, despite demonstrating that LLR is ineffective for the example of crystalline solids, our study has highlighted two useful criteria that can be applied by the research community for evaluating when best to apply LLR versus LCR.  Firstly, if the material under study has a highly irregular distribution of atoms in the interior, then LLR is very likely to offer a much better fit to the data, and therefore provide more accurate predictions for the physical quantity of interest. Secondly, LLR is also likely to give more accurate predictions in the situation where the material under study is not characterised by wide variations (discontinuities) in the predictive variable (in the example of this paper, stress) over small changes in the covariate (in this paper, distance).
Given these criteria, we believe LLR may be beneficial for similar problems in
fluid dynamics and solids with highly irregular structure (for
example, amorphous solids).
In addition, the LLR framework may also be useful for modelling 
interfacial regions between two bulk phases (considered as a
two-dimensional continuum) and the transition region near the line of
contact of three media (considered as a one-dimensional continuum).





\bibliography{stressLLR}{}
\bibliographystyle{spmpsci}      

\end{document}